# Search for neutron – mirror neutron oscillations in a laboratory experiment with ultracold neutrons

A.P. Serebrov[1*], E.B. Aleksandrov[2], N.A. Dovator[2], S.P. Dmitriev[2], A.K. Fomin[1],
P. Geltenbort[3], A.G. Kharitonov[1], I.A. Krasnoschekova[1], M.S. Lasakov[1],
A.N. Murashkin[1], G.E. Shmelev[1], V.E. Varlamov[1], A.V. Vassiljev[1],
O.M. Zherebtsov[1], O. Zimmer[3,4]

[1] *Petersburg Nuclear Physics Institute, RAS, 188300 Gatchina, Leningrad District, Russia*

[2] *Ioffe Physico-Technical Institute, RAS, 194021 St. Petersburg, Russia*

[3] *Institut Laue-Langevin, BP 156, 38042 Grenoble cedex 9, France*

[4] *Physik-Department E18, TU München, 85748 Garching, Germany*

[*] *Corresponding author*

A.P. Serebrov

Petersburg Nuclear Physics Institute

Gatchina, Leningrad district

188300 Russia

Telephone: +7 81371 46001

Fax: +7 81371 30072

E-mail: serebrov@pnpi.spb.ru

**Abstract**

Mirror matter is considered as a candidate for dark matter. In connection with this an experimental search for neutron − mirror neutron (nn′) transitions has been carried out using storage of ultracold neutrons in a trap with different magnetic fields. As a result, a new limit for the neutron − mirror neutron oscillation time $\tau_{osc}$ has been obtained, $\tau_{osc} \geq 448$ s (90% C.L.), assuming that there is no mirror magnetic field larger than 100 nT. Besides a first attempt to obtain some restriction for mirror magnetic field has been done.

*Keywords*: mirror world; neutron oscillations; ultracold neutrons

There are at least three motivations for the experimental search of mirror matter. First, in our world the weak interaction violates the P-parity and the presence of a mirror world would restore it [1]. Second, mirror matter can be considered as a natural candidate for the dark matter in the Universe [2,3]. And in third place, neutral elementary particles, e.g. photon or neutrino, could oscillate into their mirror partners [4]. In particular, it was pointed out recently [5] that a neutron – mirror neutron oscillation (nn′) could be considerably faster than neutron decay, which would have interesting experimental and astrophysical implications.

Experiments to search for nn′ transitions were carried out recently [6,7], which provided new limits on the nn′ oscillation time. Our collaboration published the best limit so far, $\tau_{osc} > 414$ s (90% C.L.) [7]. This article presents results of an additional series of experiments carried out in autumn 2007, which somewhat improve the limit. Further experiments have been performed with a wider range of magnetic fields whose implications shall be discussed as well.

The experimental setup is shown in Fig. 1. Ultracold neutrons (UCN) are trapped in a storage vessel inside a magnetic shielding which allows us to screen the Earth's magnetic field to a level below 20 nT. A solenoid within the shield can produce a homogeneous magnetic field up to a few hundred microtesla. The magnetic field settings were controlled by Cs-magnetometers.

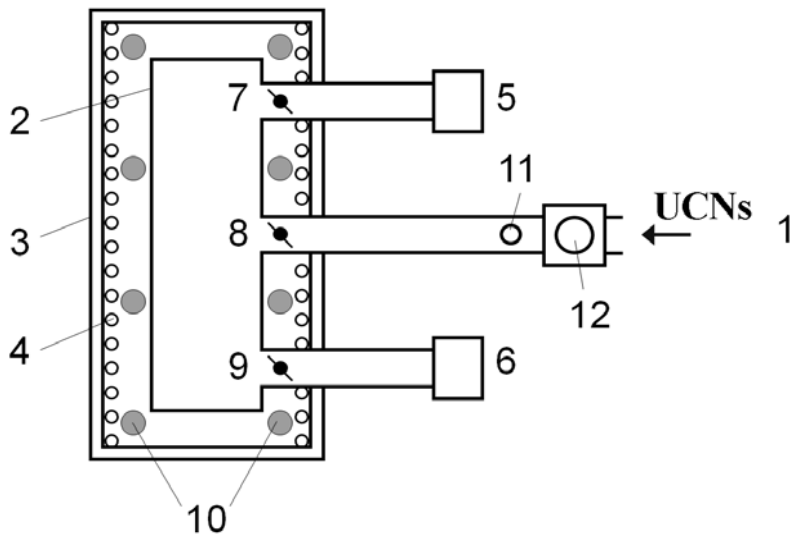

Fig. 1. Experimental setup (top view). 1: UCN input guide; 2: UCN storage chamber; 3: magnetic shielding; 4: solenoid; 5-6: UCN detectors; 7-9: valves; 10: Cs-magnetometers, 11: monitor detector, 12: entrance valve.

The experimental task is to study the dependence of the UCN storage time constant on the magnetic field. If neutron and mirror neutron have exactly the same mass and if there is an interaction mixing these states, transitions will be possible if there are no magnetic field and mirror magnetic field or if interaction with magnetic and mirror magnetic fields compensate each other. Mirror neutrons then leave the trap because they practically do not interact with ordinary matter. In the following we assume, if not stated differently and as supposed in references [6,7], the absence of a mirror magnetic field at the site of the experiment (see discussion below). A magnetic field created by the solenoid will then suppress transitions and may therefore lead to an increase of the UCN storage time constant. By measuring the numbers of neutrons in the trap $N(t_1)$ and $N(t_2)$ for a short holding time $t_1$ and a long one, $t_2$, both for magnetic field switched off, $N_0$, and switched on, $N_B$, the storage time constant is obtained. For sensitivity reasons, the long holding time $t_2$ is chosen to be close to the storage time constant of the trap itself. The searched effect will manifest as a deviation of the ratio $N_0/N_B$ from unity at $t_2$. Fig. 2 shows schematically the typical dependence of the number of neutrons after different holding times in the trap with magnetic field on or off. The effect on neutron count rates may be $10^{-3}$ or even smaller. Therefore, measurements with short holding time $t_1$ are obligatory to check for any systematic effect. For instance, the initial number of neutrons after filling the trap could depend on the magnetic field due to a small polarization of the UCN beam combined with a Stern-Gerlach effect [8]. Besides, switching on the current of the solenoid for producing the magnetic field might influence the electronic counting system. Such an effect was observed on the monitor detector used in this experiment. Although it could be suppressed by properly choosing the discriminator threshold, the monitor detector data were not used in the final data analysis. Instead, we took advantage of the high count rates in the main detectors in a continuous-flow mode, operation compared to storage mode. In the flow mode the entrance and the exit valves of the UCN trap were kept open, such that the dwell time of UCN in the trap is on average only about 20 s (calculated in a MC simulation) compared to a typical holding time of about 300 s. These measurements showed that related systematic effects are smaller than $6\times10^{-5}$.

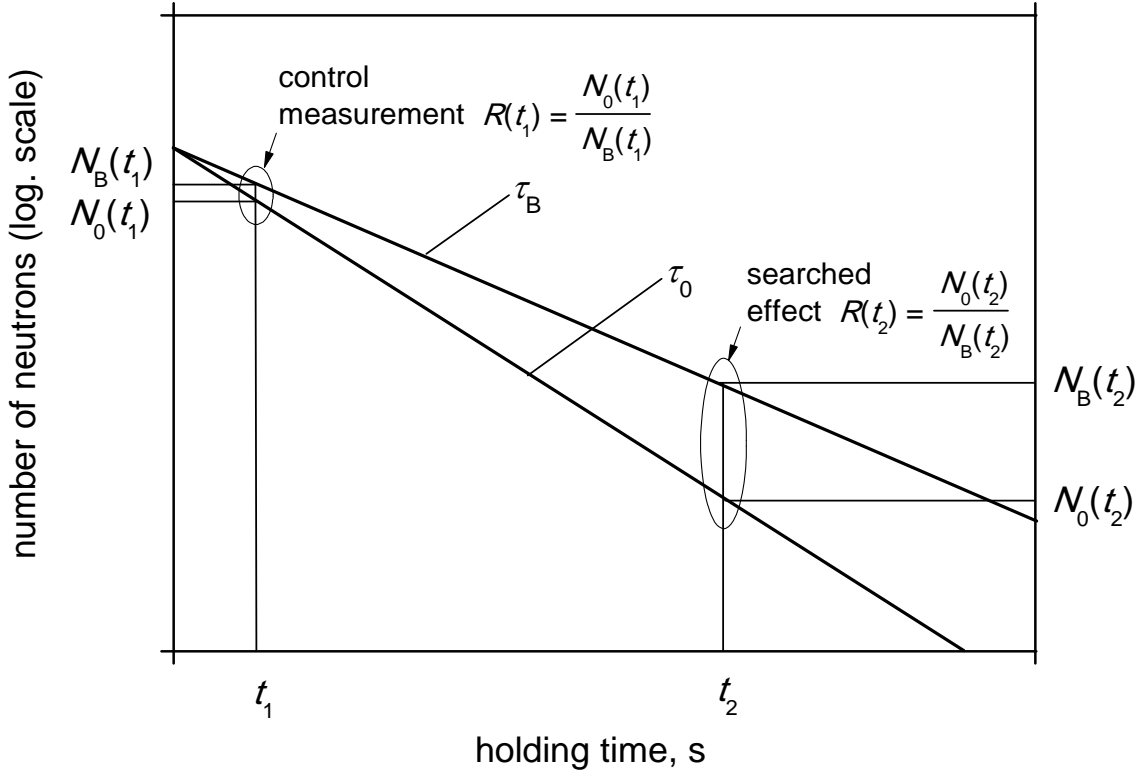

Fig. 2. Schematic representation of the exponential decrease of neutron counts in UCN storage in the trap for a holding time $t_h$. Measurements for short time $t_1$ serve for control, and the measurements with the long time $t_2$ are used to search for the effect.

The ratio of the neutron numbers $N_0/N_B$ measured at the holding time $t_2$ is related to the probability for a nn′ transition as

$$R = N_0 / N_B = \exp\left[-\bar{P}_{nn'}(t_f)\bar{n}(T_h)\right] \cong 1 - \bar{P}_{nn'}(t_f)\bar{n}(T_h). \qquad (1)$$

$\bar{P}_{nn'}(t_f)$ is the average probability for such a transition during the flight time $t_f$ of a neutron in-between collisions with the trap walls, and $\bar{n}(T_h)$ is the average number of collisions during the holding time $T_h$. For $\mu b \tau_{osc} \gg 1$ the probability for a nn′ transition is then [5,8]

$$P_{nn'}(t_f, b) = \frac{\sin^2\left[\dfrac{t_f}{\tau_{osc}}\sqrt{1+\left(\dfrac{\mu b}{2}\tau_{osc}\right)^2}\right]}{1+\left(\dfrac{\mu b}{2}\tau_{osc}\right)^2} = \frac{\sin^2\left(t_f \dfrac{\mu b}{2}\right)}{\left(\dfrac{\mu b}{2}\right)^2 \tau_{osc}^2}, \qquad (2)$$

where $\tau_{osc}$ is the oscillation time, $\mu$ the magnetic moment of the neutron, and $b$ the magnetic field. As shown in ref. [9], an exact solution of the transition probability in material traps does not add significant corrections to eq. (2), which is derived for free space without boundaries. After numerical integration over the UCN spectrum for different neutron flight times $t_f$ one obtains the final numerical dependence which can be approximated as

$$\overline{P}_{nn'}(T_h, b) = \frac{\langle t_f^2 \rangle}{\tau_{osc}^2} \overline{\nu} T_h \exp(-b/b^*). \qquad (3)$$

Here $\langle t_f^2 \rangle$ denotes the mean square time of the neutron free flight, $\overline{\nu}$ the average frequency of neutron collisions with the trap walls, $T_h$ the total holding time of neutrons in the trap (i.e. the sum of the holding time $t_h$ with the trap valves closed, the average filling time $\tau_{fill}$, and the average emptying time $\tau_{emp}$) and $b^*$ is a device specific parameter of this approximation. For the used trap – a horizontal cylinder with a diameter of 45 cm and a length of 120 cm – it is 700 nT, $\langle t_f^2 \rangle$=0.012 s$^2$ and $\overline{\nu}$=11 s$^{-1}$. Hence, a magnetic field of 700 nT will suppress the transition possibility to 1/e. More details of the setup can be found in ref. [7].

Unfortunately, there is not much information available about presence and size of mirror magnetic fields. A geophysical analysis constrains the Earth's mirror matter to below $3.8\times10^{-3}$ [10]. Model-dependent considerations of gravitational capture of dark matter bound to the Solar System estimate its total amount to $1.78\times10^{-5}$ Earth masses [11], of which only three-tenth of a percent is enclosed by the orbit of the Earth. Although there is no direct relation between mass of dark matter and mirror magnetic field, but taking the average magnetic field in our Galaxy of about 1 nT also as a "representative guess" for the mirror magnetic field, our analysis of the previous experiment [7] was performed under the assumption of a negligible mirror magnetic field. It should be noted that, however, an interaction of mirror dark matter and ordinary matter due to photon – mirror photon kinetic mixing [12] could provide an efficient mechanism to capture mirror matter in the Earth, as put forward in [13] to explain the result of the DAMA experiment to search for dark matter [14]. In this connection the papers [15,16] discuss the possible existence of mirror magnetic fields of the order of microtesla or larger. Therefore, we decided to extend our experiment to a few more magnetic field settings around zero, scanning from 20 nT up to 1200 nT, and also to increase the "strong field" setting by one order of magnitude to 20 µT. Moreover, the direction of this field was periodically changed. On one hand one achieves more statistics to improve the experimental limit for the nn′ oscillation time, adopting again the assumption that no mirror magnetic field exists and observing that this weak-field range is still sensitive to nn′ oscillation (see eq. (3)). On the other hand the presence of a mirror magnetic field in the range 0 to 1200 nT (in corresponding "mirror" units) could be verified in case an increase of the probability for nn′ oscillations will be observed due to the compensation of the mirror neutron interaction energy with the mirror magnetic field by the neutron's one with the ordinary magnetic field. No statistically significant deviation was observed. Therefore, the results of these measurements were analyzed using eq. (3) and assuming the absence of a mirror magnetic field. Fig. 3 shows the results of measurements for the dependence $\tau_{osc}^{-2} \exp(-b/b^*)$ in eq. (3) as a function of the magnetic field: $b < 20$ nT, $b = 70$ nT, 300 nT, 560 nT, and 1200 nT. From this data a new limit for nn′ oscillations can be extracted by fitting eq. (3). The result is $\tau_{osc}^{-2} = (2.84 \pm 2.03) \times 10^{-6}$ s$^{-2}$ with $\chi^2 = 1.98$, from which we derive $\tau_{osc} \geq 403$ s (90% C.L.).

Turning the argument around and supposing the existence of a nn′ mixing sufficiently large to result in a nn′ oscillation time of $\tau_{osc} \geq 200$ s (90% C.L.) for degenerate states (the weaker limit is due to the lower statistical accuracy of individual measurements), the absence of any statistically significant dependence of magnetic fields in the range 0 – 1200 nT can be interpreted as a restriction for mirror magnetic field in the same range. It should be noted that this experiment has been carried out with a horizontal direction of the magnetic field (in laboratory co-ordinates), such that the time averaged effect of the mirror magnetic field in Universe and the Solar System may be reduced by the Earth rotation. In order to obtain definite conclusions for Earth mirror magnetic field, these measurements should be carried out not only covering a much wider range of magnetic field in steps of about 400 nT, but also for three different field directions, which was not feasible within the available beam time.

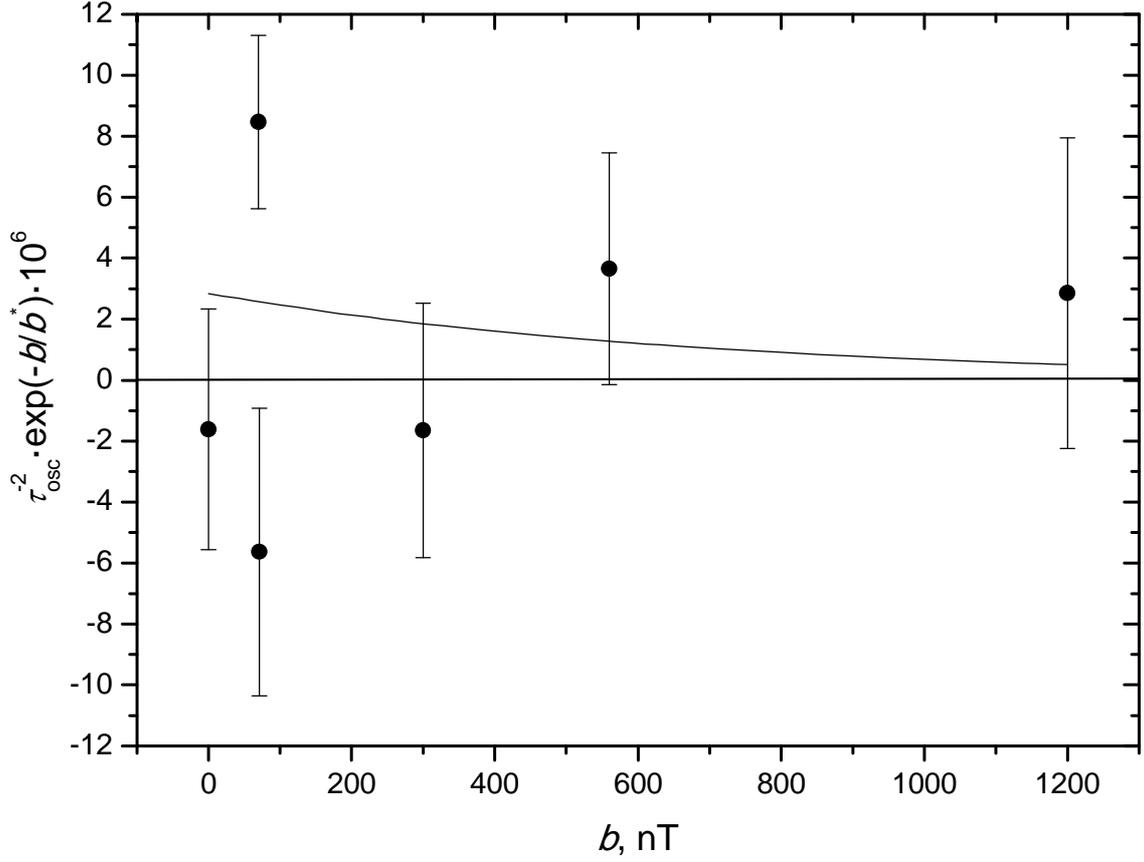

Fig. 3. Results of measurements with scanning the "zero" magnetic field. The results for $r^+ = 1 - N_{b^+}/N_{B^+}$ and $r^- = 1 - N_{b^-}/N_{B^-}$ are combined.

We obtained some additional experimental information from measurements with the "strong" magnetic field only, but in opposite directions, $B = \pm 20$ μT, in order to investigate a possible dependence on the direction of this field. The data were analyzed in terms of the ratio $R^\pm$, defined as $R^\pm = N_{B^+}/N_{B^-} = 1 - r^\pm$, which is given in Fig. 4. The result $r^\pm = (-0.06 \pm 1.01) \times 10^{-4}$ with $\chi^2 = 1.64$ indicates that there is no such dependence within the quoted accuracy. An additional series of measurements was carried out for opposite vertical magnetic fields with strength $\pm 20$ μT. The measured $r^\pm$-ratios are shown in Fig. 5. The mean value is $r^\pm = (7.5 \pm 2.4) \times 10^{-4}$ with $\chi^2 = 1.89$. To study the non-statistical dispersion of the individual results the influence of switching the current on the electronic counting system was studied using continuous-flow mode with high statistics. No effect was found on the accuracy level $10^{-4}$ (as indicated by the fourth data point before the end of the series in Fig. 5). As such control measurements were not carried out close to those points with maximum deviation 3.6σ the reason of these deviations remains unclear.

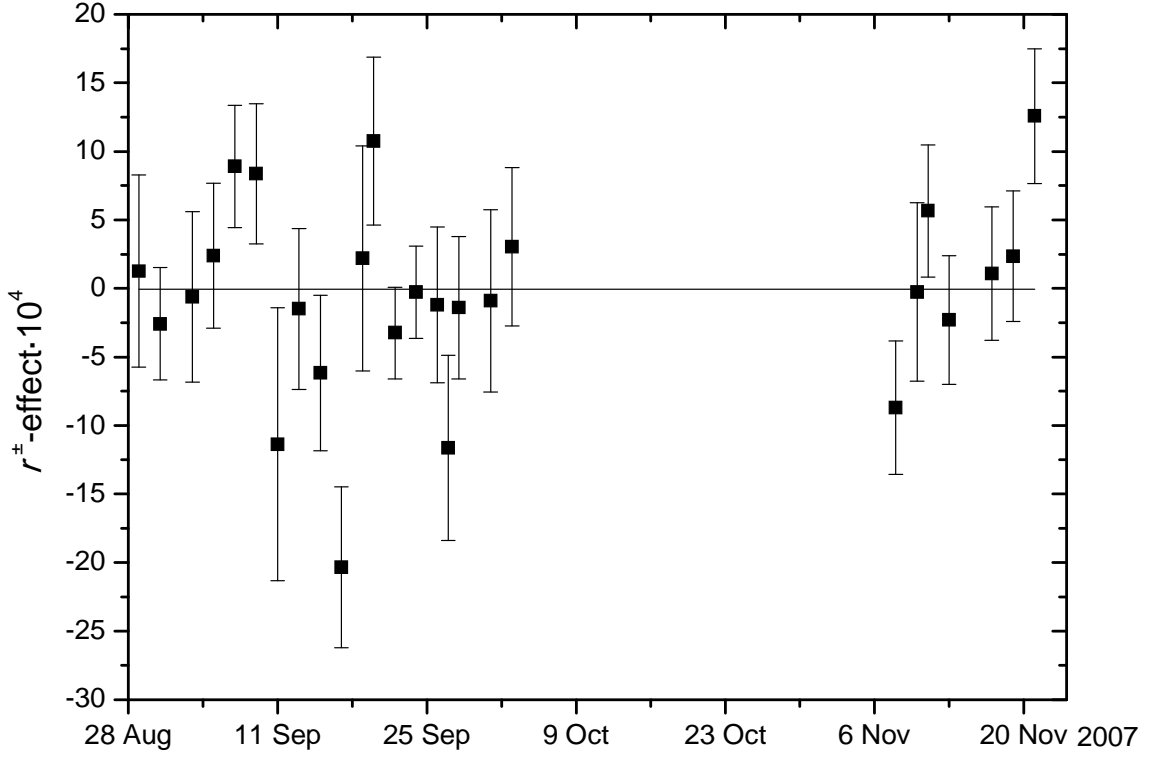

Fig. 4. Study of the effect $r^{\pm} = 1 - N_{B^+}/N_{B^-}$ with horizontal magnetic field.

To summarize, under the simplest assumption that there exists no mirror magnetic fields, we obtain $\tau_{osc}^{-2}$=(2.84 ± 2.03)×10$^{-6}$ s$^{-2}$, which corresponds to a lower limit of $\tau_{osc} \geq 403$ s (90% C.L.) on the nn′ oscillation time. Combining this result with our previous limit, $\tau_{osc}^{-2}$=(1.29 ± 2.76)×10$^{-6}$ s$^{-2}$ ($\tau_{osc} \geq 414$ s (90% C.L.)) [7], an improved limit of $\tau_{osc}^{-2}$=(2.29 ± 1.64)×10$^{-6}$ s$^{-2}$ is obtained. Hence, our improved new limit on the nn′ oscillation time is $\tau_{osc} \geq 448$ s (90% C.L.).

If one supposes the existence of a nn′ mixing sufficiently large, i.e. resulting in a nn′ oscillation time of $\tau_{osc} \geq 200$ s (90% C.L.) for degenerate states, a possible Earth mirror magnetic fields at the place of our experimental installation can be restricted in horizontal direction to 0 – 1200 nT.

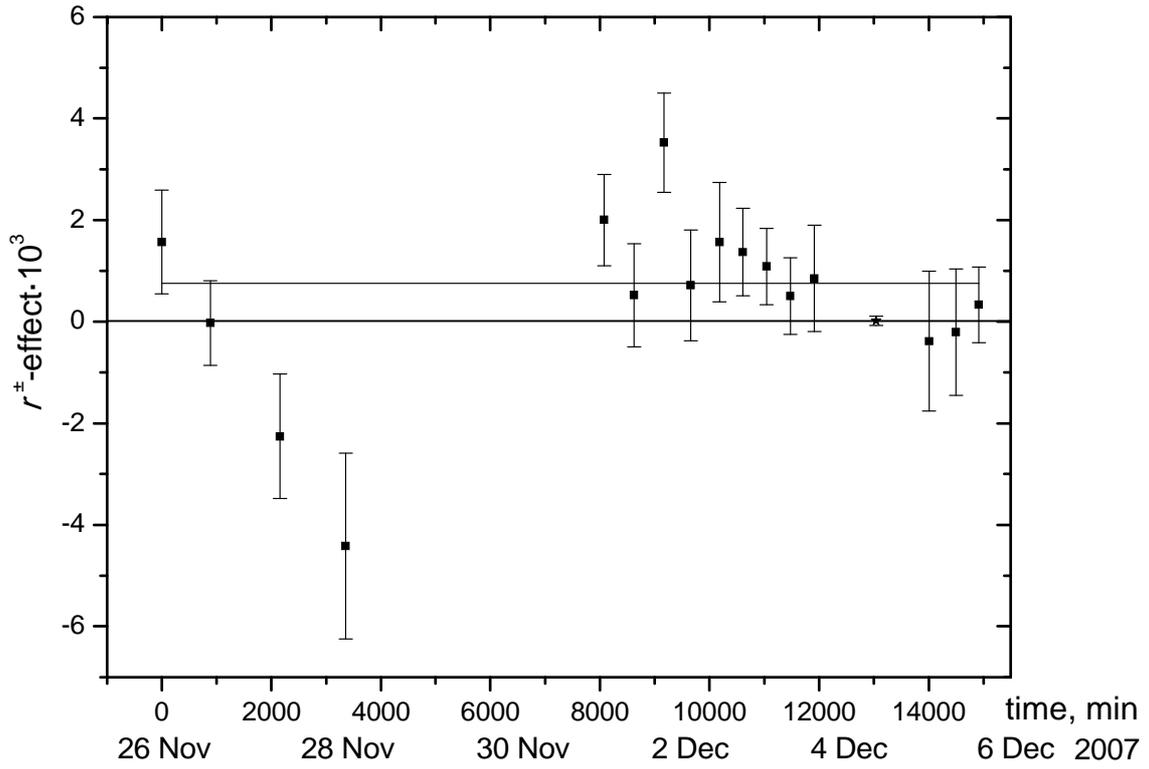

Fig. 5. Study of the effect $r^{\pm} = 1 - N_{B^+}/N_{B^-}$ with vertical magnetic field.

Acknowledgements: we would like to thank Z. Berezhiani and B. Kerbikov for useful discussions. This work has been carried out with support of the PFBR grant 07-02-00859.